\documentclass[aps,prl,twocolumn,superscriptaddress,showpacs,floatfix]{revtex4}
\usepackage{graphicx}
\usepackage{epsfig}
\begin{document}
\title{Dynamic Spin-Polarized Resonant Tunneling in Magnetic Tunnel Junctions.\\
Physical Review Letters \textbf{99}, 047206 (2007)}

\author{Casey~W.~Miller}
\altaffiliation{Present address: Physics Department, University of
South Florida, Tampa FL 33620.} \affiliation{Department of
Physics, University of California, San Diego, 9500 Gilman Drive,
La Jolla, CA 92093, USA}

\author{Zhi-Pan~Li}
\affiliation{Department of Physics, University of California, San
Diego, 9500 Gilman Drive, La Jolla, CA 92093, USA}

\author{Ivan~K.~Schuller}
\affiliation{Department of Physics, University of California, San
Diego, 9500 Gilman Drive, La Jolla, CA 92093, USA}

\author{R.~W.~Dave}
\affiliation{Technology Solutions Organization, Freescale
Semiconductor, Inc., 1300 North Alma School Road, Chandler, AZ
85224 USA}

\author{J.~M.~Slaughter}
\affiliation{Technology Solutions Organization, Freescale
Semiconductor, Inc., 1300 North Alma School Road, Chandler, AZ
85224 USA}

\author{Johan~$\rm{\AA}$kerman}
\affiliation{Department of Microelectronics and Applied Physics,
Royal Institute of Technology, Electrum 229, 164 40 Kista, Sweden}

\begin{abstract}
Precisely engineered tunnel junctions exhibit a long sought effect
that occurs when the energy of the electron is comparable to the
potential energy of the tunneling barrier. The resistance of
metal-insulator-metal tunnel junctions oscillates with an applied
voltage when electrons that tunnel directly into the barrier's
conduction band interfere upon reflection at the classical turning
points: the insulator-metal interface, and the dynamic point where
the incident electron energy equals the potential barrier inside
the insulator.  A model of tunneling between free electron bands
using the exact solution of the Schr\"{o}dinger equation for a
trapezoidal tunnel barrier qualitatively agrees with experiment.
\end{abstract}
 \maketitle

\begin{figure}
\begin{center}
\includegraphics[width=3.3in]{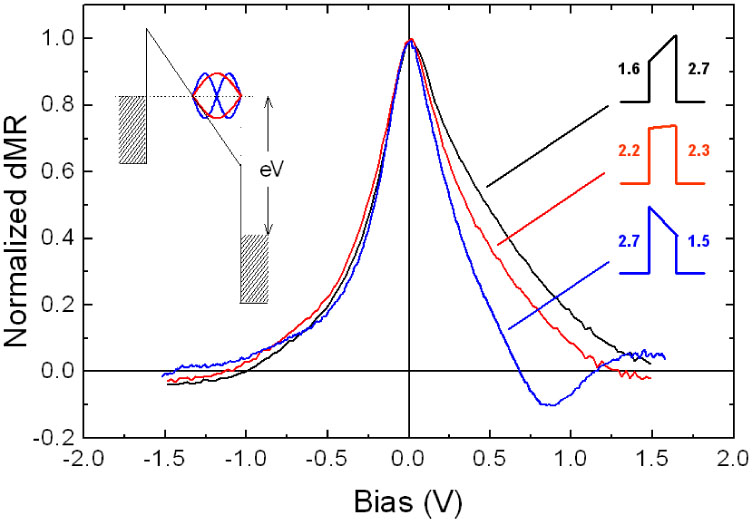}
     \caption{\small{(Color online) Normalized differential
     magnetoresistance for NiFe (bottom, blue), CFB1 (middle, red),
     and CFB2 (top, black) at 300~K. The nominal dMR(V=0) and resistance-area products
     were 80\% and 3.1~${\rm k}\Omega\cdot\mu{\rm m}^2$ for NiFe, and
     120\% and 6.0~${\rm k}\Omega\cdot\mu{\rm m}^2$ for both CFB
     devices. The barrier heights (eV) from BDR fits of dI$_p$/dV are indicated schematically to the right,
     and a cartoon of electron standing waves is shown to the left.
     }} \label{MRnorm}
\end{center}
\end{figure}

Tunneling through a barrier is one of the most fundamental
problems in physics with profound technological implications
\cite{Oppenheimer,Esaki,Giaever,JosephsonTheory,BinnigRohrer}.
Electron tunneling can be realized in multilayer systems
consisting of two conducting electrodes separated by an insulating
material. These tunneling devices can have electrodes that are
normal metals (\textit{e.g.}, Au), superconductors (Al), or
ferromagnets (Fe), while the insulators range from semiconductors
(Ge) to metal oxides (MgO). In all cases, the energy difference
between the Fermi energy of the electrodes and the conduction band
of the barrier defines the height of the tunneling barrier. The
relative energy of the tunneling electrons is varied by applying a
voltage bias between the electrodes. The barrier material's
properties generally play no role because the barrier potential is
typically much greater than the energy of the tunneling electron.
However, recent technological advances have led to robust oxide
barriers that can withstand large electric fields, which allows
access to electrons with energies comparable to the barrier
potential. It was predicted long ago that a tunneling electron
could access electronic states in the barrier and undergo resonant
tunneling \cite{gundlachSSE}, though this has not been directly
observed in metal-insulator-metal junctions with a single tunnel
barrier. This Letter reports well defined and reproducible
bias-dependent oscillations of the differential resistance in
CoFeB/MgO/NiFe tunnel junctions that are consistent with
interfering electrons within the MgO barrier. Further, we are able
to use the ferromagnetism of the electrodes to investigate these
oscillations in terms of a tunneling magnetoresistance.
Qualitative agreement with these data is obtained using a simple
tunneling model where the electrodes are treated as free electron
bands and the tunneling matrix elements are determined by solving
the Schr\"{o}dinger equation exactly for a trapezoidal barrier.\\
\indent The procedure used to fabricate our magnetic tunnel
junctions (MTJs) was reported previously \cite{intro7}. We
investigated about thirty MTJs each of CoFeB/MgO/NiFe (NiFe) and
two sets of CoFeB/MgO/CoFeB (CFB1 and CFB2), for a total of nearly
ninety junctions. As deposited CoFeB is amorphous, while NiFe is
polycrystalline. The barriers were formed by oxidizing $16\,{\rm
\AA}$ of Mg, which should produce a $13\,{\rm \AA}$ thick MgO
barrier. All devices were fabricated identically with the
exceptions of the free layer material and the procedure for
oxidation and annealing. While data presented here are for
$1~\mu{\rm m}^2$ MTJs, our observations were independent of device
shape and area, which varied from 1100~nm$\times$420~nm ellipses
to $600~{\rm nm}-10~\mu{\rm m}$ diameter circles; heating and
series resistances were thus negligible. The temperature
dependence can be attributed to thermal smearing
\cite{JohanSmear}, and all devices satisfied the MTJ tunneling
criteria \cite{MTJcriteria,MTJcriteriaJMMM}, proving
that tunneling is the primary conduction mechanism.\\
\begin{figure}
\begin{center}
\includegraphics[width=3.3in]{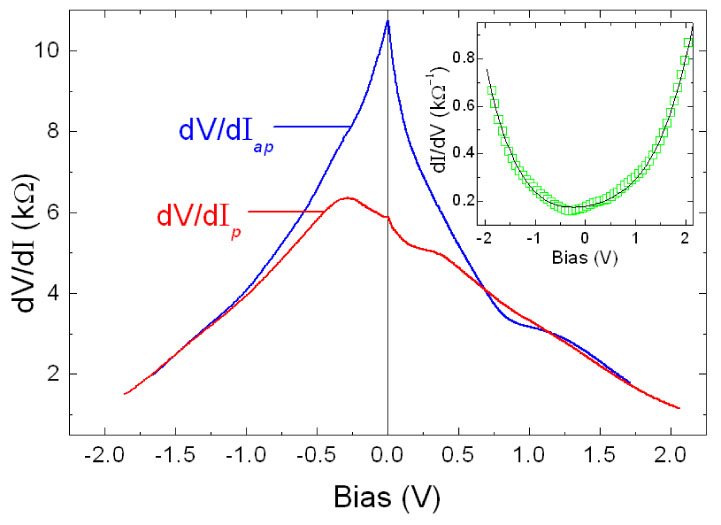}
     \caption{\small{(Color online) Differential resistance
     measurements of NiFe devices at 5~K in the $p$ and
     $ap$ magnetic configurations. (inset) The bias dependence of dI$_p$/dV for NiFe
     (green squares) is fit well by a 4$^{th}$ order polynomial (black line).}}
    \label{dVdI}
\end{center}
\end{figure}
\indent Direct measurements of the differential resistance (dV/dI)
were made with the magnetizations of the ferromagnets (in
remanence) in parallel ($p$) and antiparallel ($ap$)
configurations using a high resolution ac resistance bridge and
standard lock-in techniques. The dc bias was applied to the free
layer (NiFe or CoFeB) with the pinned layer (CoFeB) grounded.
Fig.~\ref{dVdI} shows dV/dI measurements of a NiFe device at 5~K.
An obvious oscillation for positive biases is apparent in the $ap$
state; a low amplitude oscillation is also present in the $p$
state. The dV/dI evolve continuously between these states as a
function of the angle between the free and pinned magnetizations.
dV/dI$_{p}$ equals dV/dI$_{ap}$ at several biases, the highest of
which (+1.8~V) was observable at 5\,K because of an increase in
the breakdown voltage at low temperatures. These crossing biases
were independent of temperature and the relative orientation of the ferromagnets.\\
\indent Though the underlying physics is contained in the dV/dI
for each magnetic orientation, inspecting the differential
junction magnetoresistance (dMR) is a convenient way to emphasize
this oscillatory behavior. Defining the dMR as
\vspace{-.3in}\begin{center}
\begin{eqnarray*}\label{dMR}
{\rm dMR} = \frac{{\rm dV/dI}_{ap}~-~{\rm dV/dI}_{p}}{{\rm
dV/dI}_{p}},
\end{eqnarray*}
\end{center} it is most common to observe
dMR$\,>0$ because the density of states bottleneck typically
causes ${\rm dV/dI}_{ap}$ to exceed ${\rm dV/dI}_{p}$. Thus, the
most striking feature of Fig.~\ref{MRnorm} is that the dMR of the
NiFe devices oscillates about zero when electrons tunnel from
CoFeB into NiFe. In contrast, no oscillations were observed for
either CFB1 or CFB2 (though these did have dMR$\,<0$ at high
biases). The symmetry of the biases for which dMR initially
changes sign (i.e., the first zero-crossings for both polarities)
mimics that of the barrier shape. The barrier parameters extracted
using the Brinkman, Dynes, and Rowell (BDR) model \cite{BDR} for
each device type are shown schematically in Fig.~\ref{MRnorm}.
CFB1 had a relatively symmetric barrier and symmetric
zero-crossings. The first zero-crossing for NiFe occurs when
electrons tunnel toward the low barrier interface. CFB2 had the
opposite symmetry of the NiFe devices both in barrier heights and
zero-crossings. Based on this seemingly general behavior, we
predict similar oscillations would be seen for both polarities in
NiFe/MgO/NiFe devices.
\\
\indent One possible explanation for this behavior is dynamic
resonant tunneling mediated by electron interference, a result of
Fowler-Nordheim (FN) tunneling \cite{FowlerNordheim}. In the FN
regime, the incident electron energy exceeds the potential near
the collector \textit{within the barrier}, which implies that
electrons tunnel directly into the MgO conduction band (see
cartoon in Fig.~\ref{MRnorm}). These electrons can thus be treated
as plane waves near the collector interface. Incident and
reflected electrons may then interfere and establish standing
waves within this region of the barrier (where the kinetic energy
of the electron is real). The maximum oscillation amplitude exists
in the $ap$ state, which is the result of spin-dependent
reflection from the MgO-NiFe interface. Spin-up electrons
tunneling from the CoFeB emitter with positive magnetization
toward the NiFe collector with negative magnetization are
preferentially reflected at the MgO-NiFe interface because of a
spin bottleneck in the density of states. When the applied bias
allows for resonant tunneling in the $ap$ state, spin-up emitter
electrons dominate the tunneling current. The result of this is
that the antiparallel state conductance exceeds that of the
parallel state, which leads directly to dMR$\,<0$. Negative dMR
indicates that tunneling is dominated by minority spin electrons
rather than majority electrons (as defined in the collector
electrode), and oscillations suggest that the relative conductance
of the spin species changes with bias. Gundlach showed that
oscillations of the conductance (in tunneling between normal
metals) can only be described by exactly solving the
Schr\"{o}dinger equation, and that this is evidence for the
failure of the simplest application of the
Wentzel-Kramers-Brillouin (WKB) approximation \cite{gundlachSSE}.
This effect has been identified in semiconductor systems
\cite{MOSOscillations} and scanning tunneling microscopy
\cite{STM}, but not in metal-insulator-metal
heterostructures.\\
\begin{figure}
\begin{center}
\includegraphics[width=3.3in]{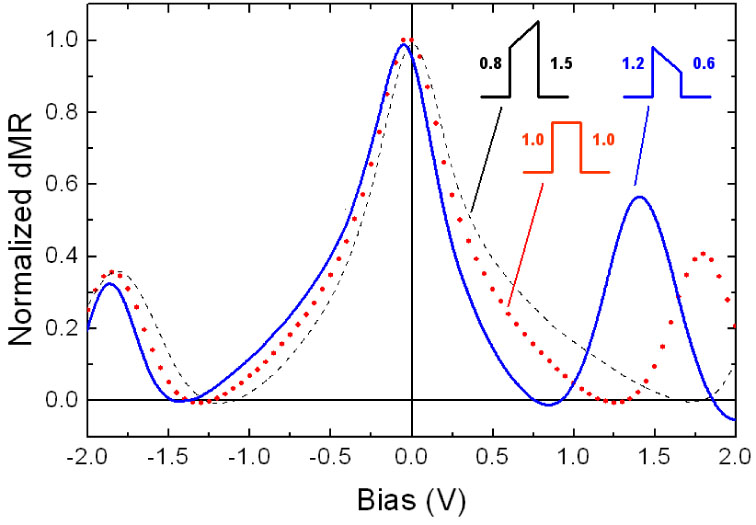}
     \caption{\small{(Color online) Model results for the dMR bias dependence for
     NiFe (blue line), CFB1 (red dots), and CFB2 (black dashes) show qualitative
     agreement with experiment. For a thickness of 28\,${\rm \AA}$ and no
     roughness, the barrier heights (eV) used in the model are indicated schematically to the right.
     }} \label{Model}
\end{center}
\end{figure}\indent Figure~\ref{Model} shows that
these experimental results are qualitatively reproduced by a model
that uses the exact solution of the Schr\"{o}dinger equation with
spin-split free electron bands representing the ferromagnetic
electrodes. A free electron model is justified here because
tunneling is dominated by $s$-like electrons \cite{Stearns,
ButlerSP,ValenzuelaTinkham}.  This model reproduces the angular
dependence of the experimental dMR when the polarization of either
electrode is varied as $\cos{\theta}$, including a nodal behavior
of the biases where dMR equals zero (i.e., when dV/dI$_{p}\equiv$
dV/dI$_{ap}$). Asymmetric barriers were approximated as
single-thickness trapezoids with barrier heights $\phi_1$ and
$\phi_3$ at the pinned and free layer interfaces. We included
material-dependent effective masses, and thus implemented the
BenDaniel-Duke boundary conditions: continuity of the wave
function $\psi_{i}(x_{{i,j}})~=~\psi_{{j}}(x_{{i,j}})$, and flux
 $\partial_{x}\psi_{i}(x_{{i,j}})/m^*_{{i}}~=~
\partial_{x}\psi_{{j}}(x_{{i,j}})/m^*_{{j}}$, at $x_{{i,j}}$
between materials $i$ and $j$ \cite{BDD}. Because \textit{ab
initio} calculations are unavailable for disordered electrodes, we
approximate the (spin-independent) effective masses
($m^*\,=\,1.3m_e$) and band bottoms (2.3~eV below the Fermi
energy, E$_F$) for both electrodes by Fe band structure
calculations \cite{DavisMacLaren}.  The MgO effective mass was
taken to be $0.4m_e$ \cite{MgOmass}. The spin splittings
($\Delta$) of the electrodes were estimated via $\Delta_i~=~{\rm
E}_F[(1-P_i)^2/(1+P_i)^2-1]$, where $i$ denotes the material, and
the polarizations ($P$) of CoFeB and NiFe were taken to be 55\%
and 45\% from tunneling spin polarization measurements
\cite{ParkinCoFeMgO,ParkinNiFe}. The tunneling currents in the
parallel and antiparallel states were calculated including the
spin-dependent densities of states \cite{XiaoDOS} via
\vspace{-.3in}\begin{center}
 \begin{eqnarray*}\label{Pcurrent}
{\rm I}_{p}\propto\int\,\left({\rm
D}_{\uparrow\uparrow}N_\uparrow^{(1)}N_\uparrow^{(3)}+{\rm
D}_{\downarrow\downarrow}N_\downarrow^{(1)}N_\downarrow^{(3)}\right)F(E)\,{\rm
dE},\nonumber\\
{\rm I}_{ap}\propto\int\,\left({\rm
D}_{\uparrow\downarrow}N_\uparrow^{(1)}N_\downarrow^{(3)}+{\rm
D}_{\downarrow\uparrow}N_\downarrow^{(1)}N_\uparrow^{(3)}\right)F(E)\,{\rm
dE},
\end{eqnarray*}
\end{center}
\noindent where subscripts denote the pertinent spin sub-bands,
superscripts denote the materials, and $F(E)$ is the difference in
Fermi functions of the two electrodes $[
f^{(1)}(E)-f^{(3)}(E-e{\rm V})]$. The spin-dependent densities of
states for materials 1 (grounded, pinned layer) and 3 (biased free
layer) are respectively $N_m^{(1)}=N_m^{(1)}(E)$ and
$N_n^{(3)}=N_n^{(3)}(E-e{\rm V})$, where subscripts represent the
spin sub-bands.  The tunneling matrix elements D$_{mn}={\rm
D}_{mn}(s, \phi_1, \phi_3, {\rm V}, E)$ are the probabilities of
tunneling between spin sub-bands $m$ and $n$ in electrodes 1 and
3, respectively, and were obtained by exactly solving the
Schr\"{o}dinger equation. Spin-flip processes were neglected. The
upper limit of the transverse integration was truncated at 2\% of
the Fermi energy because the tunneling current is dominated by
wave vectors within a cone of $\sim8^{\circ}$ from normal
incidence \cite{DynesDirection}.  This model is simple and
therefore appealing, though more intricate analyses
\cite{MathonUmerskiSpacer} may capture more effectively the
underlying spin-dependent physics.\\
\indent The location and amplitude of the negative dMR peak shift
only slightly when interfacial roughness is included in the
calculation in a previously demonstrated manner
\cite{MillerRoughness}. This is because the bias of the initial
oscillation is set by the barrier height at the collector
interface (the threshold bias for FN tunneling equals this
height). Note that for CFB1 or CFB2 oscillations were not observed
because the barrier heights were comparable to the maximum applied
bias.  On the other hand, a more significant effect is seen for
the positive peak around 1.4\,V. Including 15\% roughness, which
was the roughness determined for similar MTJs \cite{ParabolicBad},
causes the amplitude of the latter peak to fall from 58\% to 45\%,
and requires a mean thickness of 35\,${\rm \AA}$ to keep the dMR
peaks at biases consistent with the data of Fig.\,\ref{MRnorm};
the discrepancy from the growth thickness is
reasonable for this qualitative model.\\
\indent The model needed to explain the present data implies
barrier parameters that are different from those obtained by BDR
fits. To qualitatively reproduce the data, the model requires
similar $\phi_1$ with only $\phi_3$ significantly different
between the three MTJ types (shown schematically in
Fig.~\ref{Model}). This is reasonable since the fabrication
process for all the devices was identical until the barrier
oxidation; perturbations due to this step should predominately
affect $\phi_3$, not $\phi_1$. The BDR fits, on the other hand,
indicate a roughly constant average barrier height with both
$\phi_1$ and $\phi_3$ different for each MTJ type, which is less
reasonable considering the fabrication procedure.  The model
calculations use a thickness of 28~${\rm \AA}$ (when neglecting
roughness), while the BDR fits yield $\sim8~{\rm \AA}$. This
discrepancy is most likely due to interfacial roughness and the
WKB approximation failing in these devices (as indicated by
the electron interference presented here).\\
\indent FN tunneling is required for the interference of electrons
within the barrier. In this regime, both the width and average
height of the barrier at the Fermi energy decrease with increasing
bias. The result of this should be an exponentially increasing
conductance for applied biases greater than the collector
interface barrier height. The parallel state conductance data are
fit well by a fourth-order polynomial (Fig.~\ref{dVdI} inset),
showing that deviations from the low-bias parabolic behavior
exist, but the data are not exponentially increasing at the
highest biases. The delayed transition to exponential conductance
may be related to the effective mass of the tunneling electron,
interfacial roughness, or may be an emerging characteristic of
coherent tunneling through crystalline barriers,
possibly originating from heating effects \cite{ParabolicBad, NateMgB2}.\\
\indent Similar oscillations are obtained in resonant tunneling
studies where a normal metal spacer exists between the barrier and
one ferromagnet (see e.g., Ref.~\cite{YuasaResTunnel} and
references therein).  We refer to those as ``static" phenomena
because the thickness of the normal metal is fixed. The phenomenon
we report here is ``dynamic" because the thickness of the
interference region can be tuned by the applied bias: electrons
inside the MgO conduction band can be treated as free electrons,
making this region directly analogous to the normal metal spacer
used in static studies. Additionally, the present case allows the
effective mass of the MgO conduction band to be estimated from the
oscillation period. Recalling that the de\,Broglie wavelength (in
${\rm \AA}$) is $\lambda\,=\,0.529\varepsilon m_e/m^*$, where
$\varepsilon$ is the dielectric constant ($\varepsilon\sim3$) and
$m^*/m_e$ is the reduced mass of MgO \cite{MgObandgap}, the
measured oscillation period corresponds to $\lambda\sim2.6\,{\rm
\AA}$, and $m^*\sim0.6m_e$, in reasonable agreement with
the expected 0.4$m_e$ \cite{MgOmass}.\\
\indent Alternate explanations for our observations can be ruled
out. The oscillations are odd functions of bias, and thus cannot
be explained by emission phenomena such as magnons
\cite{RowellChapter}. The persistence of the oscillatory nature at
finite bias with interface roughness excludes localized barrier
states and interface resonant states \cite{TsymbalResInversion,
tsymbal:401}. The angular dependence is strong evidence that this
effect is not due to quantum size effects in the electrodes
\cite{JaklevicStandingWaves}. A similar negative dMR region in
Co$_2$MnSi/MgO/CoFe tunnel junctions was recently interpreted as
an energy gap in the Co$_2$MnSi minority-spin band
\cite{YamamotoHeuslerOsc}, but this explanation does not apply to
our case because no such gap is expected for NiFe.  While it is
not possible to rule out unknown density of states effects,
dynamic resonant tunneling is the most convincing origin because
of its simplicity and ability to reproduce numerous experimental
features from different MTJs with different materials.\\
\indent  In summary, novel bias-dependent oscillations were
observed in the differential resistances of metal-insulator-metal
(CoFeB/MgO/NiFe) tunnel junctions. These long sought oscillations
are due to dynamic resonant tunneling mediated by interference of
electrons that tunnel into the conduction band of the insulator
(MgO). A coherent tunneling model using the exact solution of the
Schr\"{o}dinger equation and free electrons representing the
electrodes qualitatively reproduced the bias dependence.  The
tunability of this newly demonstrated phenomenon, as well as its
spin-dependence, may help advance the development of tunable
resonant tunneling systems for fundamental spintronics physics and
applications.
\begin{acknowledgments}
Supported by the US DOE, the Swedish Foundation for Strategic
Research, The Swedish Research Council, and The G\"{o}ran
Gustafsson Foundation. The authors thank J. M. Rowell, R. C.
Dynes, H. Suhl, L. J. Sham, H. Dery and L. Cywinski for useful
discussions, and D. Mix for help with the wafer level
measurements.
\end{acknowledgments}

%
\end{document}